\documentclass[superscriptaddress,preprintnumbers,amsmath,amssymb,twocolumn,floats,aps,prl,reprint]{revtex4-1}
\usepackage{txfonts}
\usepackage{amssymb}
\usepackage{graphicx}
\usepackage{sidecap}
\usepackage{CJK}
\usepackage{txfonts}
\usepackage{url}
\usepackage[]{sidecap}
\usepackage{color}

\begin{document}

\bibliographystyle{unsrt}
\title{Impurity-pinned incommensurate charge density wave and local phonon excitations in $\textbf{2\emph{H}}$-NbS$_2$}
\author{Chenhaoping Wen}
\thanks{These two authors contributed equally.}
\author{Yuan Xie}
\thanks{These two authors contributed equally.}
\author{Yueshen Wu}
\author{Shiwei Shen}
\author{Pengfei Kong}
\affiliation{School of Physical Science and Technology, ShanghaiTech University, Shanghai 201210, China}
\author{Hailong Lian}
\affiliation{Key Laboratory of Artificial Structures and Quantum Control and Shanghai Center for Complex Physics, Department of Physics and Astronomy, Shanghai Jiao Tong University, Shanghai 200240, China}
\author{Jun Li}
\affiliation{School of Physical Science and Technology, ShanghaiTech University, Shanghai 201210, China}
\affiliation{ShanghaiTech Laboratory for Topological Physics, ShanghaiTech University, Shanghai 201210, China}
\author{Hui Xing}
\affiliation{Key Laboratory of Artificial Structures and Quantum Control and Shanghai Center for Complex Physics, Department of Physics and Astronomy, Shanghai Jiao Tong University, Shanghai 200240, China}
\author{Shichao Yan}
\email{yanshch@shanghaitech.edu.cn}
\affiliation{School of Physical Science and Technology, ShanghaiTech University, Shanghai 201210, China}
\affiliation{ShanghaiTech Laboratory for Topological Physics, ShanghaiTech University, Shanghai 201210, China}
\date{\today}


\begin{abstract}
 Here we report a scanning tunneling microscopy (STM) and spectroscopy (STS) study in the superconducting state of 2$H$-NbS$_2$. We directly visualize the existence of incommensurate charge density wave (CDW) that is pinned by atomic impurities. In strong tunneling conditions, the incommensurate CDW is de-pinned from impurities by the electric field from STM tip. We perform STM-based inelastic tunneling spectroscopy (IETS) to detect phonon excitations in 2$H$-NbS$_2$ and measure the influence of atomic impurities on local phonon excitations. In comparison with the calculated vibrational density of states in 2$H$-NbS$_2$, we find two branches of phonon excitations which correspond to the vibrations of Nb ions and S ions, and the strength of the local phonon excitations is insensitive to the atomic impurities. Our results demonstrate the coexistence of incommensurate CDW and superconductivity in $2H$-NbS$_2$, and open the way of detecting atomic-scale phonon excitations in transition metal dichalcogenides with STM-based IETS.
\end{abstract}

\maketitle
Electron-phonon and electron-electron interactions in transition metal dichalcogenides (TMDs) produce a wide range of collective electronic states, such as charge density wave, superconductivity, and Mott-insulating phases, which makes TMDs a fertile playground for exploring the interplay between these novel electronic phases~\cite{Tutis_2008_NM, Cava_2006_NP, Rossnagel_2011_JPCM, Reznik_2011_PRL, Vidya_2017_PRL, Cossu_2020_NPG}. For the superconducting TMDs, one of the common features is that superconductivity coexists or emerges in proximity to a charge density wave (CDW) state~\cite{Tutis_2008_NM, Cava_2006_NP}. In the family of superconducting TMDs, $2H$-NbS$_2$ stands out as it is often considered as a superconducting TMDs material without CDW~\cite{Tanaka_1983_JPSJ, Rodiere_2008_PRL}. It is even more surprising if we consider $2H$-NbS$_2$ has similar superconducting transition temperature as its isostructural and isoelectronic compound, $2H$-NbSe$_2$, where 3 $\times$ 3 CDW order coexists with superconductivity~\cite{Soumyanarayanan_2013_PNAS}.

\begin{figure}[htb]
\includegraphics[width=86mm]{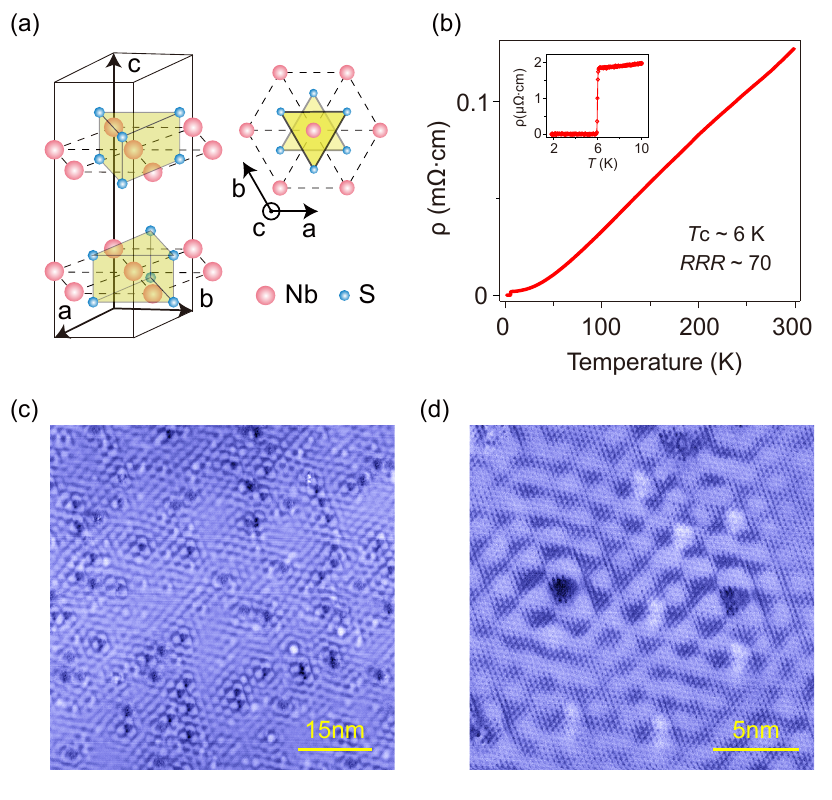}
\caption{
(a) Schematic crystal structure of $2H$-NbS$_2$. (b) Temperature dependent electrical resistivity for 2$H$-NbS$_2$. The inset shows the resistivity data near superconducting transition. The residual resistance ratio (\textit{RRR}) is about 70. (c) and (d) STM constant current topographies taken with $V_s$ = 500~mV, $I$ = 10~pA.
}
\label{Crystal}
\end{figure}

In $2H$-NbS$_2$, no clear experimental evidence of CDW phase has been reported in electrical transport~\cite{Tanaka_1983_JPSJ}, low-temperature scanning tunneling microscopy (STM)~\cite{Rodiere_2008_PRL} and high-pressure measurements~\cite{Suderow_2013_PRB}. Until recently, the diffuse x-ray scattering measurements report faint traces of $\sqrt{13} \times\sqrt{13}$ CDW order in $2H$-NbS$_2$, where they suspect this may be due to the presence of rare and dilute $1T$ layers in their $2H$-NbS$_2$ samples~\cite{Rodiere_2018_PRB}. The \emph{k}-dependent electron-phonon coupling is often considered to be essential for the CDW origin in TMDs, and the absence of CDW in $2H$-NbS$_2$ is theoretically explained as the anharmonic effects that suppress the formation of CDW, although there are phonon modes softening~\cite{Rodiere_2012_PRB, Francesco_2019_NanoLett}.

\begin{figure*}[htb]
\centering
\includegraphics[width=145mm]{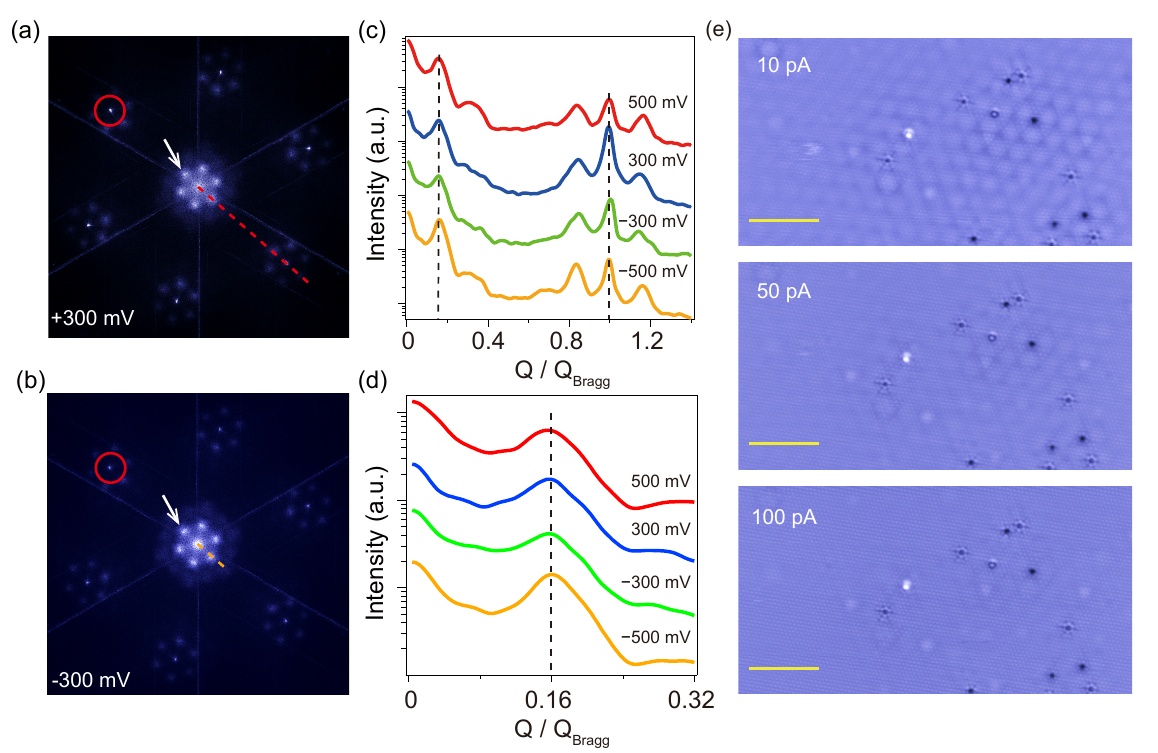}
\caption{
(a)-(b) FT of STM images taken on the same region as Fig.~\ref{Crystal} (c) with +300~mV and -300~mV bias voltages, respectively. The FT images are symmetrized along the high-symmetry directions. The red circles are the Bragg peaks and the white arrows indicate the CDW peaks. (c) Line-cut profiles taken along the red dashed line shown in panel (a) for the FT of STM topographies with different STM bias voltages (curves are shifted for clarity). (d)  Line-cut profiles as in panel (c), but along a shorter orange dashed line shown in panel (b). (e) A sequence of STM topographies on the same area with different tunneling current setpoints: 10~pA (top), 50~pA (middle), 100~pA (bottom) and the same bias voltage $V_s$ = 2~V. Scale bar: 5.6 nm.
}
\label{CDW}
\end{figure*}

In the CDW phase of real materials, the total energy of the system is minimized with charge modulation either by locking in some special way with atomic lattices (commensurate CDW case), or by 'pinning' to atomic impurities (incommensurate CDW case). Impurities always exist even in high-quality single crystals, and impurity-scattering has a profound effect in the static and dynamic behaviors of CDW~\cite{Lee_1978_PRB, Akbari_2018_PRB}. For example, impurities in materials can be the pinning centers of short-range CDWs and the seeds for the formation of long-range CDWs~\cite{Schmidt_2016_ACSNano, Pasupathy_2014_PRB, Millis_2015_PRB}. However, in the previous theoretical calculations explaining the absence of CDW in $2H$-NbS$_2$, only perfect $2H$-NbS$_2$ structure was considered~\cite{Rodiere_2012_PRB, Francesco_2019_NanoLett}. And in the low-temperature STM studies on $2H$-NbS$_2$, only STM images taken on small defect free regions were shown~\cite{Rodiere_2008_PRL}. To clearly prove whether CDW exists in the superconducting state of $2H$-NbS$_2$ with atomic impurities or not, detailed low-temperature local probe investigations are still missing.

\begin{figure}[htb]
\centering
\includegraphics[width=86mm]{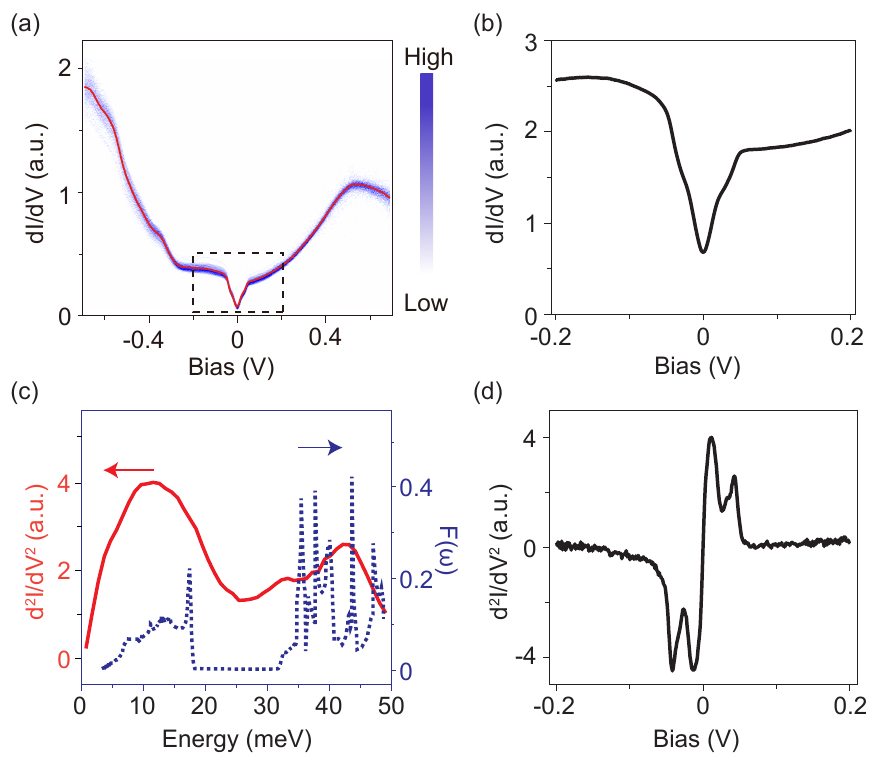}
\caption{
(a) The distribution of the spatially resolved d$I$/d$V$ spectra (blue) acquired over a 20 $\times$ 20~nm$^2$ region, and the averaged d$I$/d$V$ spectrum (red). (b) Averaged d$I$/d$V$ spectrum in a smaller energy range. (c) The positive bias side of the d$^2I$/d$V^2$ spectrum in panel (d) (red), and the previously calculated vibrational density of states in $2H$-NbS$_2$ (dotted blue) [31]. (d) Averaged d$^2I$/d$V^2$ spectrum taken at the same time with the d$I$/d$V$ spectrum in panel (b).
}
\label{STS}
\end{figure}

In this letter, we use low-temperature scanning tunneling microscopy (STM) and spectroscopy (STS) to study the high-quality $2H$-NbS$_2$ single crystals. We discover that incommensurate CDW exists in the superconducting state of $2H$-NbS$_2$, and the incommensurate CDW patterns are pinned by the atomic impurities. Since electron-phonon coupling is essential for the charge instability in many TMDs materials, we use inelastic tunneling spectroscopy (IETS) to locally probe phonon excitations and investigate the influence of atomic impurities on phonon excitations in $2H$-NbS$_2$.


Single crystals of $2H$-NbS$_2$ are synthesized by chemical vapor transport method as described elsewhere~\cite{Liu_2017_PhysicaC}. Fig.~\ref{Crystal} (b) shows the electrical resistivity as a function of temperature for the $2H$-NbS$_2$ single crystal sample used in this work. The smooth temperature dependence in the resistance is consistent with the previously reported data~\cite{Liu_2017_PhysicaC}, and the measured sharp superconducting transition temperature is at $\sim$6~K ~\cite{SuppleRef} (also see Supplementary Material, S1). STM experiments were carried out with a Unisoku low-temperature STM at the base temperature of $\sim$4.5~K (below the superconducting transition temperature of $2H$-NbS$_2$). $2H$-NbS$_2$ single crystal samples were cleaved at 77~K and then transferred into STM head for measurements. Chemically etched tungsten tips were used for the measurements. STS measurements were done by using standard lock-in technique with 3 or 5~mV modulation at the frequency of 437~Hz.

For the structure of $2H$-NbS$_2$, each unit consists of two S-Nb-S sandwich layers in which the S sheets have a hexagonal close-packed structure and the Nb atoms are in the trigonal prismatic coordination defined by the two S sheets [Fig.~\ref{Crystal} (a)]. The sample cleaves between S-Nb-S sandwich layers terminated in a S surface. Constant current STM topographies [Figs.~\ref{Crystal} (c) and (d)] show three clear features: (1) The surface S atoms form triangular lattice; (2) The surface and subsurface impurities, such as vacancies or substitution atoms, are clearly visible, and their densities are low and comparable with typical high-quality TMDs single crystals~\cite{Pasupathy_2014_PRB}; (3)  CDW patterns can be clearly seen in the vicinity of the atomic impurities. While the intensity gradually decreases as far away from the impurities, the CDW patterns cover almost the entire surface. As shown in Figs.~\ref{Crystal} (c) and (d), the CDW phase is pinned at each impurity which indicates they are in the strong-pinning case~\cite{Littlewood_1981_PRB}.

In order to reveal the periodicity of the CDW order, we perform Fourier transform (FT) to STM topographies measured on the same region as shown in Fig.~\ref{Crystal} (c). Figs.~\ref{CDW} (a) and (b) shows FT images of STM topographies taken with +500~mV and -500~mV bias voltages, respectively.  Both FT images show the hexagonal Bragg peaks (red circles) and CDW vectors (indicated by white arrows). Bragg peaks appear as sharp bright spots while CDW peaks are slightly broader and closer to the center of FT images. Bragg peaks and CDW vectors are almost along the same direction. Figs.~\ref{CDW} (c) and (d) show the line-cut profiles along Bragg peak direction in FT images taken with +500~mV, +300~mV, -300~mV and -500~mV bias voltages~\cite{SuppleRef} (see Supplementary Material, S2). We can clearly see that the CDW vector is bias voltage independent, and it is located at $\sim$0.16$\cdot$Q$_{Bragg}$, which indicates the CDW order is incommensurate with the atomic S lattice.

Now a natural question is whether this impurity-pinned patterns in Figs.~\ref{CDW} (a) and (b) are not CDW, but are due to the quasiparticle interface (QPI) induced electronic standing waves near impurities~\cite{Eigler_1993_Nature, Monceau_2012_PRL}. To address this question, we perform current dependent measurements while keeping STM bias voltage constant. If the patterns are electronic standing waves, as increasing the tunneling current, signal-to-noise in STM topography will be improved and the electronic standing waves will become clearer. In our current dependent measurements [Fig.~\ref{CDW} (e)], as increasing tunneling current, the impurity-pinned patterns become weaker and gradually disappear. They recover as we decrease the strength of the tunneling current~\cite{SuppleRef} (see Supplemental Material, S3). This excludes the possibility that this impurity-pinned patterns are the electronic standing waves, and it is consistent with the effect that the CDW patterns are de-pinned by the electric field from the STM tip in strong tunneling conditions~\cite{Rice_1979_PRB}. In the STM topographies taken with strong tunneling conditions, individual impurities can be clearly resolved, where we can see that CDW minimum is pinned at the S-vacancy-like impurities, and CDW maximum is pinned at the S-substitution-atoms~\cite{SuppleRef} (see Supplementary Material, S4). Our measurements indicate that the impurity-pinned CDW is fragile and can be disturbed by the electric field from STM tip, which may also explain why it has not been found in the previous low-temperature STM measurements ~\cite{Rodiere_2008_PRL}.

After confirming the CDW nature of the impurity-pinned patterns, we perform STS measurements to get information about their possible mechanism. For low-dimensional CDW systems, Fermi surface nesting and \emph{k}-dependent electron-phonon coupling are often considered as the driving forces for the CDW instability~\cite{Rossnagel_2011_JPCM, Mazin_2008_PRB}. \emph{k}-dependent electron-phonon coupling is essential for CDW formation in many TMDs~\cite{Rossnagel_2011_JPCM}, and STM-based inelastic tunneling spectroscopy (STM-IETS) has been used to detect phonon excitations in Pb thin films and graphene~\cite{Wulfhekel_2015_PRL, Komeda_2017_PRB, Crommie_2008_NP}. STM-IETS signal is due to the opening of an inelastic tunneling channel as the energy of the tunneling electrons is larger than the threshold of the excitation, which increases the differential tunneling conductance, d$I$/d$V$. STM-IETS signal is usually weak and detected by recording the second derivative of the tunneling current as a function of STM bias voltage (d$^2I$/d$V^2$) with lock-in amplifier~\cite{Ho_1998_Science}.

Fig.~\ref{STS} (a) shows the position-dependent d$I$/d$V$ spectra and averaged d$I$/d$V$ spectrum in the energy range of $\pm$700~mV. The strength of the d$I$/d$V$ signal is proportional to the local density of states (LDOS). As shown in Fig.~\ref{STS} (a), d$I$/d$V$ spectrum on $2H$-NbS$_2$ is spatially uniform, and there are no clear impurity-induced resonant states. According to the calculated electronic band structure of $2H$-NbS$_2$~\cite{Giustino_2018_PRB}, the peak at $\sim$+500~mV in Fig.~\ref{STS} (a) is likely related to the top of the band with out-of-plane Nb d$_{z^2}$ character. No clear QPI pattern is observed in the d$I$/d$V$ maps over a large energy range~\cite{SuppleRef} (see Supplementary Material, S5), which confirms again that the patterns near impurities shown in Fig.~\ref{Crystal} (b) are not electronic standing waves. Another feature in the d$I$/d$V$ spectrum is that there is a pronounced dip near Fermi level, and Fig.~\ref{STS} (b) shows the high-resolution d$I$/d$V$ spectrum in a smaller energy range near Fermi level. Although the experimental temperature ($\sim$4.5~K) is below the superconducting transition temperature of $2H$-NbS$_2$ ($\sim$6~K), due to thermal broadening, the superconducting gap of $2H$-NbS$_2$ cannot be clearly detected in the d$I$/d$V$ spectrum at 4.5~K~\cite{Rodiere_2008_PRL}. As shown in Fig.~\ref{STS} (b), there are two steps on both positive and negative sides of the Fermi level. Fig.~\ref{STS} (d) is the simultaneously recorded d$^2I$/d$V^2$ spectrum as the d$I$/d$V$ spectrum in Fig.~\ref{STS} (b), where we can see two clear features: (1) it is more or less symmetric about Fermi level; (2) there are two peak-dip pairs at bias voltages of $\pm$12~mV and $\pm$42~mV. These two features are the characteristic fingerprints of inelastic tunneling processes~\cite{Wulfhekel_2015_PRL, Ho_1998_Science}. We note that the same d$^2I$/d$V^2$ spectrum can be obtained with another fresh STM tip which confirms that the d$^2I$/d$V^2$ spectrum is STM tip independent~\cite{SuppleRef} (Supplementary Material, S6).

In order to understand the IETS processes, we compare the measured d$^2I$/d$V^2$ spectrum with previously calculated vibrational density of states (DOS) in $2H$-NbS$_2$~\cite{Motizuki_1994_JPSJ}. As shown in Fig.~\ref{STS} (c), they are consistent with each other, which indicates the STM-IETS spectrum is directly related to phonon excitations in $2H$-NbS$_2$. According to the calculated vibrational DOS, there are two branches of vibrations, which correspond to the vibration of Nb ions (with energies around 12~mV) and S ions (with energies around 42~mV), respectively. Due to the limited energy resolution at the measured temperature (4.5~K), more detailed features in the calculated vibrational DOS cannot be resolved in the d$^2I$/d$V^2$ spectrum [Fig.~\ref{STS} (c)]. Identifying the dip-like feature near Fermi level in d$I$/d$V$ spectrum as the inelastic phonon excitations will also be helpful for understanding similar features in the d$I$/d$V$ spectra taken on the other TMDs, such as $2H$-NbSe$_2$ and $3R$-NbS$_2$~\cite{Soumyanarayanan_2013_PNAS, Machida_2017_PRB}.

\begin{figure}[htb]
\includegraphics[width=86mm]{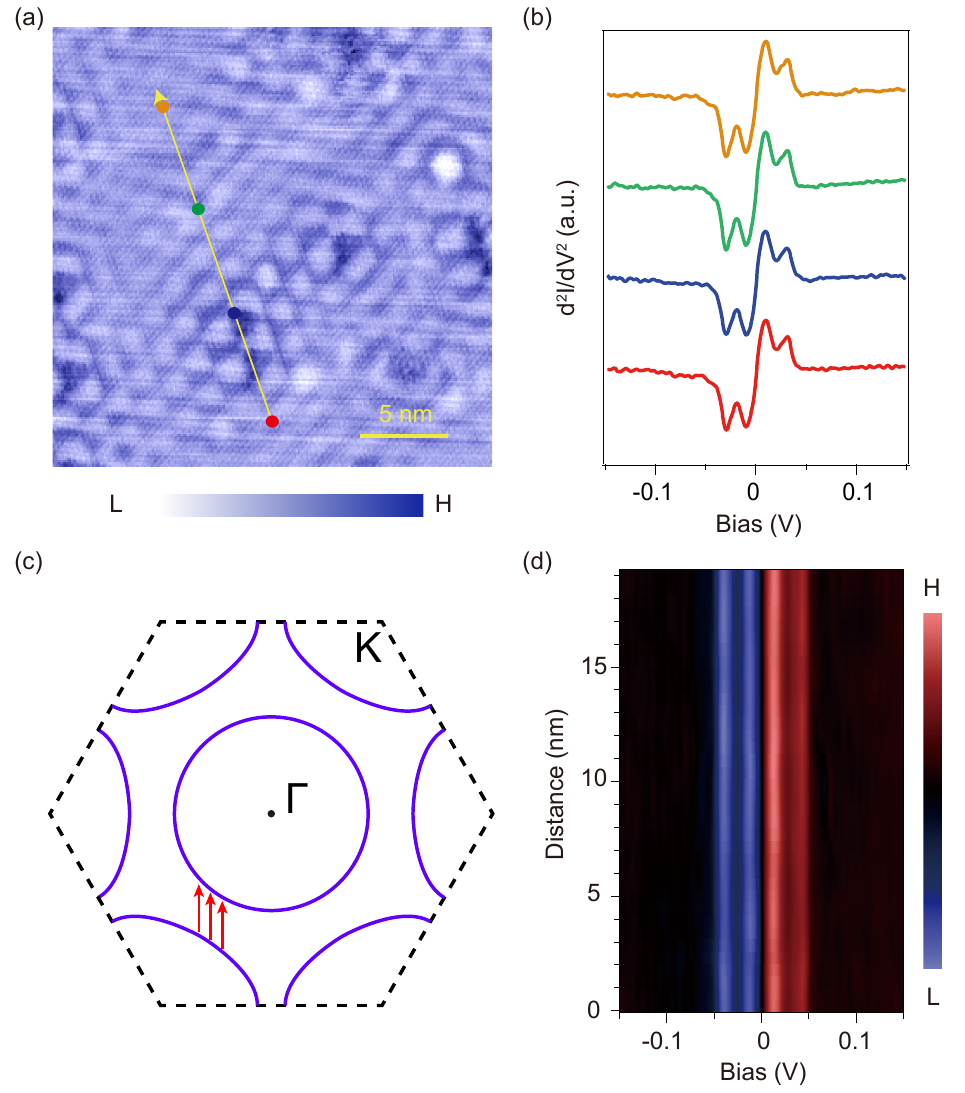}
\caption{
(a) STM constant current topography taken with $V_s$ = 500~mV, $I$ = 10~pA. (b) d$^2I$/d$V^2$ spectra taken on the four colored dots in panel (a). (c) Schematic Fermi surface of $2H$-NbS$_2$ based on ARPES data in~\cite{Sirica_2016_PRB}. The dashed hexagon is the Brillouin zone and the red arrows indicate the nesting vectors. (d) d$^2I$/d$V^2$ spectra as a function of distance (vertical axis) taken along the yellow line in panel (a).
}
\label{Linecut}
\end{figure}

Since the incommensurate CDW in $2H$-NbS$_2$ is pinned by the atomic impurities and electron-phonon coupling is often considered as the driving force for the charge instability in TMDs, we then measure the influence of the atomic impurities on the local phonon excitations. As shown in Figs.~\ref{Linecut} (a) and (b), the STM-IETS spectra taken at the positions near and far away from the impurity are more or less the same. Fig.~\ref{Linecut} (d) shows the line-cut STM-IETS spectra along the line shown in Fig.~\ref{Linecut} (a), and we can see that the energy positions and the strength of the local phonon excitations are quite uniform~\cite{SuppleRef} (Supplementary Material, S7). This means that although the incommensurate CDW is pinned by the atomic impurities, the local phonon excitations are impurity insensitive, which indicates that electron-phonon coupling may not be the dominant factor for the formation of incommensurate CDW in $2H$-NbS$_2$. Otherwise, we would expect to detect changes in local phonon excitations near the atomic impurities. Based on our STM-IETS data, we propose that the most likely mechanism of the impurity-pinned incommensurate CDW in $2H$-NbS$_2$ is Fermi surface nesting. Fig.~\ref{Linecut} (c) shows the schematic Fermi surface of $2H$-NbS$_2$ \cite{Sirica_2016_PRB}. As can be seen, the hole pockets are located at $\Gamma$ and K points, and there are parallel ('nested') sections in them. The red arrow indicates the possible nesting vector which is comparable with the incommensurate CDW vector shown in Fig.~\ref{CDW} (d). We also find that in some $2H$-NbS$_2$ samples, the intensity of the impurity-pinned incommensurate CDW is weaker, and this could be due to the weakened Fermi surface nesting caused by the slightly different doping levels in those samples.


In Summary, we discovered impurity-pinned incommensurate CDW in the superconducting state of $2H$-NbS$_2$, which indicates the incommensurate CDW coexists with superconductivity in $2H$-NbS$_2$. This discovery will be helpful for unifying the picture of the interplay between CDW and superconductivity in TMDs. We also demonstrated that STM-IETS could be used to detect local phonon excitations in $2H$-NbS$_2$, which opens the door for detecting spatially-resolved phonon excitations in TMDs. Further STM/STS studies at even lower temperature (below 1~K) would be helpful to measure more detailed information about the phonon excitations with better energy resolution, and it would also be important to study the relationship between local phonon excitations and the superconducting properties of $2H$-NbS$_2$.


We thank professors Fu-Chun Zhang, Xi Chen and Xinghua Lu for valuable discussions. S.Y. acknowledges the financial support from Science and Technology Commission of Shanghai Municipality (STCSM) (Grant No. 18QA1403100), National Science Foundation of China (Grant No. 11874042) and the start-up funding from ShanghaiTech University. J.L. acknowledges the financial support from National Science Foundation of China (Grant No. 61771234).

\textit{Note added.} During the reversion of the manuscript, we became aware of related work about STM based IETS on TMDs ~\cite{IETS_PRL}.

\bibliographystyle{apsrev4-1}

\begin{thebibliography}{34}%
\makeatletter
\providecommand \@ifxundefined [1]{%
 \@ifx{#1\undefined}
}%
\providecommand \@ifnum [1]{%
 \ifnum #1\expandafter \@firstoftwo
 \else \expandafter \@secondoftwo
 \fi
}%
\providecommand \@ifx [1]{%
 \ifx #1\expandafter \@firstoftwo
 \else \expandafter \@secondoftwo
 \fi
}%
\providecommand \natexlab [1]{#1}%
\providecommand \enquote  [1]{``#1''}%
\providecommand \bibnamefont  [1]{#1}%
\providecommand \bibfnamefont [1]{#1}%
\providecommand \citenamefont [1]{#1}%
\providecommand \href@noop [0]{\@secondoftwo}%
\providecommand \href [0]{\begingroup \@sanitize@url \@href}%
\providecommand \@href[1]{\@@startlink{#1}\@@href}%
\providecommand \@@href[1]{\endgroup#1\@@endlink}%
\providecommand \@sanitize@url [0]{\catcode `\\12\catcode `\$12\catcode
  `\&12\catcode `\#12\catcode `\^12\catcode `\_12\catcode `\%12\relax}%
\providecommand \@@startlink[1]{}%
\providecommand \@@endlink[0]{}%
\providecommand \url  [0]{\begingroup\@sanitize@url \@url }%
\providecommand \@url [1]{\endgroup\@href {#1}{\urlprefix }}%
\providecommand \urlprefix  [0]{URL }%
\providecommand \Eprint [0]{\href }%
\providecommand \doibase [0]{http://dx.doi.org/}%
\providecommand \selectlanguage [0]{\@gobble}%
\providecommand \bibinfo  [0]{\@secondoftwo}%
\providecommand \bibfield  [0]{\@secondoftwo}%
\providecommand \translation [1]{[#1]}%
\providecommand \BibitemOpen [0]{}%
\providecommand \bibitemStop [0]{}%
\providecommand \bibitemNoStop [0]{.\EOS\space}%
\providecommand \EOS [0]{\spacefactor3000\relax}%
\providecommand \BibitemShut  [1]{\csname bibitem#1\endcsname}%
\let\auto@bib@innerbib\@empty
\bibitem [{\citenamefont {Sipos}\ \emph {et~al.}(2008)\citenamefont {Sipos},
  \citenamefont {Kusmartseva}, \citenamefont {Akrap}, \citenamefont {Berger},
  \citenamefont {Forr\'{o}},\ and\ \citenamefont {Tuti\v{s}}}]{Tutis_2008_NM}%
  \BibitemOpen
  \bibfield  {author} {\bibinfo {author} {\bibfnamefont {B.}~\bibnamefont
  {Sipos}}, \bibinfo {author} {\bibfnamefont {A.~F.}\ \bibnamefont
  {Kusmartseva}}, \bibinfo {author} {\bibfnamefont {A.}~\bibnamefont {Akrap}},
  \bibinfo {author} {\bibfnamefont {H.}~\bibnamefont {Berger}}, \bibinfo
  {author} {\bibfnamefont {L.}~\bibnamefont {Forr\'{o}}}, \ and\ \bibinfo
  {author} {\bibfnamefont {E.}~\bibnamefont {Tuti\v{s}}},\ }\href {\doibase
  10.1038/nmat2318} {\bibfield  {journal} {\bibinfo  {journal} {Nature
  Materials}\ }\textbf {\bibinfo {volume} {7}},\ \bibinfo {pages} {960}
  (\bibinfo {year} {2008})}\BibitemShut {NoStop}%
\bibitem [{\citenamefont {Morosan}\ \emph {et~al.}(2006)\citenamefont
  {Morosan}, \citenamefont {Zandbergen}, \citenamefont {Dennis}, \citenamefont
  {Bos}, \citenamefont {Onose}, \citenamefont {Klimczuk}, \citenamefont
  {Ramirez}, \citenamefont {Ong},\ and\ \citenamefont {Cava}}]{Cava_2006_NP}%
  \BibitemOpen
  \bibfield  {author} {\bibinfo {author} {\bibfnamefont {E.}~\bibnamefont
  {Morosan}}, \bibinfo {author} {\bibfnamefont {H.~W.}\ \bibnamefont
  {Zandbergen}}, \bibinfo {author} {\bibfnamefont {B.~S.}\ \bibnamefont
  {Dennis}}, \bibinfo {author} {\bibfnamefont {J.~W.~G.}\ \bibnamefont {Bos}},
  \bibinfo {author} {\bibfnamefont {Y.}~\bibnamefont {Onose}}, \bibinfo
  {author} {\bibfnamefont {T.}~\bibnamefont {Klimczuk}}, \bibinfo {author}
  {\bibfnamefont {A.~P.}\ \bibnamefont {Ramirez}}, \bibinfo {author}
  {\bibfnamefont {N.~P.}\ \bibnamefont {Ong}}, \ and\ \bibinfo {author}
  {\bibfnamefont {R.~J.}\ \bibnamefont {Cava}},\ }\href {\doibase
  10.1038/nphys360} {\bibfield  {journal} {\bibinfo  {journal} {Nature
  Physics}\ }\textbf {\bibinfo {volume} {2}},\ \bibinfo {pages} {544} (\bibinfo
  {year} {2006})}\BibitemShut {NoStop}%
\bibitem [{\citenamefont {Rossnagel}(2011)}]{Rossnagel_2011_JPCM}%
  \BibitemOpen
  \bibfield  {author} {\bibinfo {author} {\bibfnamefont {K.}~\bibnamefont
  {Rossnagel}},\ }\href {\doibase 10.1088/0953-8984/23/21/213001} {\bibfield
  {journal} {\bibinfo  {journal} {Journal of Physics: Condensed Matter}\
  }\textbf {\bibinfo {volume} {23}},\ \bibinfo {pages} {213001} (\bibinfo
  {year} {2011})}\BibitemShut {NoStop}%
\bibitem [{\citenamefont {Weber}\ \emph {et~al.}(2011)\citenamefont {Weber},
  \citenamefont {Rosenkranz}, \citenamefont {Castellan}, \citenamefont
  {Osborn}, \citenamefont {Hott}, \citenamefont {Heid}, \citenamefont {Bohnen},
  \citenamefont {Egami}, \citenamefont {Said},\ and\ \citenamefont
  {Reznik}}]{Reznik_2011_PRL}%
  \BibitemOpen
  \bibfield  {author} {\bibinfo {author} {\bibfnamefont {F.}~\bibnamefont
  {Weber}}, \bibinfo {author} {\bibfnamefont {S.}~\bibnamefont {Rosenkranz}},
  \bibinfo {author} {\bibfnamefont {J.-P.}\ \bibnamefont {Castellan}}, \bibinfo
  {author} {\bibfnamefont {R.}~\bibnamefont {Osborn}}, \bibinfo {author}
  {\bibfnamefont {R.}~\bibnamefont {Hott}}, \bibinfo {author} {\bibfnamefont
  {R.}~\bibnamefont {Heid}}, \bibinfo {author} {\bibfnamefont {K.-P.}\
  \bibnamefont {Bohnen}}, \bibinfo {author} {\bibfnamefont {T.}~\bibnamefont
  {Egami}}, \bibinfo {author} {\bibfnamefont {A.~H.}\ \bibnamefont {Said}}, \
  and\ \bibinfo {author} {\bibfnamefont {D.}~\bibnamefont {Reznik}},\ }\href
  {\doibase 10.1103/PhysRevLett.107.107403} {\bibfield  {journal} {\bibinfo
  {journal} {Physical Review Letters}\ }\textbf {\bibinfo {volume} {107}},\
  \bibinfo {pages} {107403} (\bibinfo {year} {2011})}\BibitemShut {NoStop}%
\bibitem [{\citenamefont {Yan}\ \emph {et~al.}(2017)\citenamefont {Yan},
  \citenamefont {Iaia}, \citenamefont {Morosan}, \citenamefont {Fradkin},
  \citenamefont {Abbamonte},\ and\ \citenamefont {Madhavan}}]{Vidya_2017_PRL}%
  \BibitemOpen
  \bibfield  {author} {\bibinfo {author} {\bibfnamefont {S.}~\bibnamefont
  {Yan}}, \bibinfo {author} {\bibfnamefont {D.}~\bibnamefont {Iaia}}, \bibinfo
  {author} {\bibfnamefont {E.}~\bibnamefont {Morosan}}, \bibinfo {author}
  {\bibfnamefont {E.}~\bibnamefont {Fradkin}}, \bibinfo {author} {\bibfnamefont
  {P.}~\bibnamefont {Abbamonte}}, \ and\ \bibinfo {author} {\bibfnamefont
  {V.}~\bibnamefont {Madhavan}},\ }\href {\doibase
  10.1103/PhysRevLett.118.106405} {\bibfield  {journal} {\bibinfo  {journal}
  {Physical Review Letters}\ }\textbf {\bibinfo {volume} {118}},\ \bibinfo
  {pages} {106405} (\bibinfo {year} {2017})}\BibitemShut {NoStop}%
\bibitem [{\citenamefont {Cossu}\ \emph {et~al.}(2020)\citenamefont {Cossu},
  \citenamefont {Palot{\'a}s}, \citenamefont {Sarkar}, \citenamefont
  {Di~Marco},\ and\ \citenamefont {Akbari}}]{Cossu_2020_NPG}%
  \BibitemOpen
  \bibfield  {author} {\bibinfo {author} {\bibfnamefont {F.}~\bibnamefont
  {Cossu}}, \bibinfo {author} {\bibfnamefont {K.}~\bibnamefont {Palot{\'a}s}},
  \bibinfo {author} {\bibfnamefont {S.}~\bibnamefont {Sarkar}}, \bibinfo
  {author} {\bibfnamefont {I.}~\bibnamefont {Di~Marco}}, \ and\ \bibinfo
  {author} {\bibfnamefont {A.}~\bibnamefont {Akbari}},\ }\href {\doibase
  10.1038/s41427-020-0207-x} {\bibfield  {journal} {\bibinfo  {journal} {NPG
  Asia Materials}\ }\textbf {\bibinfo {volume} {12}},\ \bibinfo {pages} {24}
  (\bibinfo {year} {2020})}\BibitemShut {NoStop}%
\bibitem [{\citenamefont {Naito}\ and\ \citenamefont
  {Tanaka}(1982)}]{Tanaka_1983_JPSJ}%
  \BibitemOpen
  \bibfield  {author} {\bibinfo {author} {\bibfnamefont {M.}~\bibnamefont
  {Naito}}\ and\ \bibinfo {author} {\bibfnamefont {S.}~\bibnamefont {Tanaka}},\
  }\href {\doibase 10.1143/JPSJ.51.219} {\bibfield  {journal} {\bibinfo
  {journal} {Journal of the Physical Society of Japan}\ }\textbf {\bibinfo
  {volume} {51}},\ \bibinfo {pages} {219} (\bibinfo {year} {1982})}\BibitemShut
  {NoStop}%
\bibitem [{\citenamefont {Guillam\'{o}n}\ \emph {et~al.}(2008)\citenamefont
  {Guillam\'{o}n}, \citenamefont {Suderow}, \citenamefont {Vieira},
  \citenamefont {Cario}, \citenamefont {Diener},\ and\ \citenamefont
  {Rodi\.{e}re}}]{Rodiere_2008_PRL}%
  \BibitemOpen
  \bibfield  {author} {\bibinfo {author} {\bibfnamefont {I.}~\bibnamefont
  {Guillam\'{o}n}}, \bibinfo {author} {\bibfnamefont {H.}~\bibnamefont
  {Suderow}}, \bibinfo {author} {\bibfnamefont {S.}~\bibnamefont {Vieira}},
  \bibinfo {author} {\bibfnamefont {L.}~\bibnamefont {Cario}}, \bibinfo
  {author} {\bibfnamefont {P.}~\bibnamefont {Diener}}, \ and\ \bibinfo {author}
  {\bibfnamefont {P.}~\bibnamefont {Rodi\.{e}re}},\ }\href {\doibase
  10.1103/PhysRevLett.101.166407} {\bibfield  {journal} {\bibinfo  {journal}
  {Physical Review Letters}\ }\textbf {\bibinfo {volume} {101}},\ \bibinfo
  {pages} {166407} (\bibinfo {year} {2008})}\BibitemShut {NoStop}%
\bibitem [{\citenamefont {Soumyanarayanan}\ \emph {et~al.}(2013)\citenamefont
  {Soumyanarayanan}, \citenamefont {Yee}, \citenamefont {He}, \citenamefont
  {van Wezel}, \citenamefont {Rahn}, \citenamefont {Rossnagel}, \citenamefont
  {Hudson}, \citenamefont {Norman},\ and\ \citenamefont
  {Hoffman}}]{Soumyanarayanan_2013_PNAS}%
  \BibitemOpen
  \bibfield  {author} {\bibinfo {author} {\bibfnamefont {A.}~\bibnamefont
  {Soumyanarayanan}}, \bibinfo {author} {\bibfnamefont {M.~M.}\ \bibnamefont
  {Yee}}, \bibinfo {author} {\bibfnamefont {Y.}~\bibnamefont {He}}, \bibinfo
  {author} {\bibfnamefont {J.}~\bibnamefont {van Wezel}}, \bibinfo {author}
  {\bibfnamefont {D.~J.}\ \bibnamefont {Rahn}}, \bibinfo {author}
  {\bibfnamefont {K.}~\bibnamefont {Rossnagel}}, \bibinfo {author}
  {\bibfnamefont {E.~W.}\ \bibnamefont {Hudson}}, \bibinfo {author}
  {\bibfnamefont {M.~R.}\ \bibnamefont {Norman}}, \ and\ \bibinfo {author}
  {\bibfnamefont {J.~E.}\ \bibnamefont {Hoffman}},\ }\href {\doibase
  10.1073/pnas.1211387110} {\bibfield  {journal} {\bibinfo  {journal}
  {Proceedings of the National Academy of Sciences}\ }\textbf {\bibinfo
  {volume} {110}},\ \bibinfo {pages} {1623} (\bibinfo {year}
  {2013})}\BibitemShut {NoStop}%
\bibitem [{\citenamefont {Tissen}\ \emph {et~al.}(2013)\citenamefont {Tissen},
  \citenamefont {Osorio}, \citenamefont {Brison}, \citenamefont {Nemes},
  \citenamefont {Garc\'{i}a-Hern\'{a}ndez}, \citenamefont {Cario},
  \citenamefont {Rodi\`ere}, \citenamefont {Vieira},\ and\ \citenamefont
  {Suderow}}]{Suderow_2013_PRB}%
  \BibitemOpen
  \bibfield  {author} {\bibinfo {author} {\bibfnamefont {V.~G.}\ \bibnamefont
  {Tissen}}, \bibinfo {author} {\bibfnamefont {M.~R.}\ \bibnamefont {Osorio}},
  \bibinfo {author} {\bibfnamefont {J.~P.}\ \bibnamefont {Brison}}, \bibinfo
  {author} {\bibfnamefont {N.~M.}\ \bibnamefont {Nemes}}, \bibinfo {author}
  {\bibfnamefont {M.}~\bibnamefont {Garc\'{i}a-Hern\'{a}ndez}}, \bibinfo
  {author} {\bibfnamefont {L.}~\bibnamefont {Cario}}, \bibinfo {author}
  {\bibfnamefont {P.}~\bibnamefont {Rodi\`ere}}, \bibinfo {author}
  {\bibfnamefont {S.}~\bibnamefont {Vieira}}, \ and\ \bibinfo {author}
  {\bibfnamefont {H.}~\bibnamefont {Suderow}},\ }\href {\doibase
  10.1103/PhysRevB.87.134502} {\bibfield  {journal} {\bibinfo  {journal}
  {Physical Review B}\ }\textbf {\bibinfo {volume} {87}},\ \bibinfo {pages}
  {134502} (\bibinfo {year} {2013})}\BibitemShut {NoStop}%
\bibitem [{\citenamefont {Leroux}\ \emph {et~al.}(2018)\citenamefont {Leroux},
  \citenamefont {Cario}, \citenamefont {Bosak},\ and\ \citenamefont
  {Rodi\'{e}re}}]{Rodiere_2018_PRB}%
  \BibitemOpen
  \bibfield  {author} {\bibinfo {author} {\bibfnamefont {M.}~\bibnamefont
  {Leroux}}, \bibinfo {author} {\bibfnamefont {L.}~\bibnamefont {Cario}},
  \bibinfo {author} {\bibfnamefont {A.}~\bibnamefont {Bosak}}, \ and\ \bibinfo
  {author} {\bibfnamefont {P.}~\bibnamefont {Rodi\'{e}re}},\ }\href {\doibase
  10.1103/PhysRevB.97.195140} {\bibfield  {journal} {\bibinfo  {journal}
  {Physical Review B}\ }\textbf {\bibinfo {volume} {97}},\ \bibinfo {pages}
  {195140} (\bibinfo {year} {2018})}\BibitemShut {NoStop}%
\bibitem [{\citenamefont {Leroux}\ \emph {et~al.}(2012)\citenamefont {Leroux},
  \citenamefont {Le~Tacon}, \citenamefont {Calandra}, \citenamefont {Cario},
  \citenamefont {M\'easson}, \citenamefont {Diener}, \citenamefont
  {Borrissenko}, \citenamefont {Bosak},\ and\ \citenamefont
  {Rodi\`ere}}]{Rodiere_2012_PRB}%
  \BibitemOpen
  \bibfield  {author} {\bibinfo {author} {\bibfnamefont {M.}~\bibnamefont
  {Leroux}}, \bibinfo {author} {\bibfnamefont {M.}~\bibnamefont {Le~Tacon}},
  \bibinfo {author} {\bibfnamefont {M.}~\bibnamefont {Calandra}}, \bibinfo
  {author} {\bibfnamefont {L.}~\bibnamefont {Cario}}, \bibinfo {author}
  {\bibfnamefont {M.-A.}\ \bibnamefont {M\'easson}}, \bibinfo {author}
  {\bibfnamefont {P.}~\bibnamefont {Diener}}, \bibinfo {author} {\bibfnamefont
  {E.}~\bibnamefont {Borrissenko}}, \bibinfo {author} {\bibfnamefont
  {A.}~\bibnamefont {Bosak}}, \ and\ \bibinfo {author} {\bibfnamefont
  {P.}~\bibnamefont {Rodi\`ere}},\ }\href {\doibase 10.1103/PhysRevB.86.155125}
  {\bibfield  {journal} {\bibinfo  {journal} {Physical Review B}\ }\textbf
  {\bibinfo {volume} {86}},\ \bibinfo {pages} {155125} (\bibinfo {year}
  {2012})}\BibitemShut {NoStop}%
\bibitem [{\citenamefont {Bianco}\ \emph {et~al.}(2019)\citenamefont {Bianco},
  \citenamefont {Errea}, \citenamefont {Monacelli}, \citenamefont {Calandra},\
  and\ \citenamefont {Mauri}}]{Francesco_2019_NanoLett}%
  \BibitemOpen
  \bibfield  {author} {\bibinfo {author} {\bibfnamefont {R.}~\bibnamefont
  {Bianco}}, \bibinfo {author} {\bibfnamefont {I.}~\bibnamefont {Errea}},
  \bibinfo {author} {\bibfnamefont {L.}~\bibnamefont {Monacelli}}, \bibinfo
  {author} {\bibfnamefont {M.}~\bibnamefont {Calandra}}, \ and\ \bibinfo
  {author} {\bibfnamefont {F.}~\bibnamefont {Mauri}},\ }\href {\doibase
  10.1021/acs.nanolett.9b00504} {\bibfield  {journal} {\bibinfo  {journal}
  {Nano Letters}\ }\textbf {\bibinfo {volume} {19}},\ \bibinfo {pages} {3098}
  (\bibinfo {year} {2019})}\BibitemShut {NoStop}%
\bibitem [{\citenamefont {Fukuyama}\ and\ \citenamefont
  {Lee}(1978)}]{Lee_1978_PRB}%
  \BibitemOpen
  \bibfield  {author} {\bibinfo {author} {\bibfnamefont {H.}~\bibnamefont
  {Fukuyama}}\ and\ \bibinfo {author} {\bibfnamefont {P.~A.}\ \bibnamefont
  {Lee}},\ }\href {\doibase 10.1103/PhysRevB.17.535} {\bibfield  {journal}
  {\bibinfo  {journal} {Physical Review B}\ }\textbf {\bibinfo {volume} {17}},\
  \bibinfo {pages} {535} (\bibinfo {year} {1978})}\BibitemShut {NoStop}%
\bibitem [{\citenamefont {Cossu}\ \emph {et~al.}(2018)\citenamefont {Cossu},
  \citenamefont {Moghaddam}, \citenamefont {Kim}, \citenamefont {Tahini},
  \citenamefont {Di~Marco}, \citenamefont {Yeom},\ and\ \citenamefont
  {Akbari}}]{Akbari_2018_PRB}%
  \BibitemOpen
  \bibfield  {author} {\bibinfo {author} {\bibfnamefont {F.}~\bibnamefont
  {Cossu}}, \bibinfo {author} {\bibfnamefont {A.~G.}\ \bibnamefont
  {Moghaddam}}, \bibinfo {author} {\bibfnamefont {K.}~\bibnamefont {Kim}},
  \bibinfo {author} {\bibfnamefont {H.~A.}\ \bibnamefont {Tahini}}, \bibinfo
  {author} {\bibfnamefont {I.}~\bibnamefont {Di~Marco}}, \bibinfo {author}
  {\bibfnamefont {H.-W.}\ \bibnamefont {Yeom}}, \ and\ \bibinfo {author}
  {\bibfnamefont {A.}~\bibnamefont {Akbari}},\ }\href {\doibase
  10.1103/PhysRevB.98.195419} {\bibfield  {journal} {\bibinfo  {journal} {Phys.
  Rev. B}\ }\textbf {\bibinfo {volume} {98}},\ \bibinfo {pages} {195419}
  (\bibinfo {year} {2018})}\BibitemShut {NoStop}%
\bibitem [{\citenamefont {Yeom}\ \emph {et~al.}(2016)\citenamefont {Yeom},
  \citenamefont {Oh}, \citenamefont {Wippermann},\ and\ \citenamefont
  {Schmidt}}]{Schmidt_2016_ACSNano}%
  \BibitemOpen
  \bibfield  {author} {\bibinfo {author} {\bibfnamefont {H.~W.}\ \bibnamefont
  {Yeom}}, \bibinfo {author} {\bibfnamefont {D.~M.}\ \bibnamefont {Oh}},
  \bibinfo {author} {\bibfnamefont {S.}~\bibnamefont {Wippermann}}, \ and\
  \bibinfo {author} {\bibfnamefont {W.~G.}\ \bibnamefont {Schmidt}},\ }\href
  {\doibase 10.1021/acsnano.5b05925} {\bibfield  {journal} {\bibinfo  {journal}
  {ACS Nano}\ }\textbf {\bibinfo {volume} {10}},\ \bibinfo {pages} {810}
  (\bibinfo {year} {2016})}\BibitemShut {NoStop}%
\bibitem [{\citenamefont {Arguello}\ \emph {et~al.}(2014)\citenamefont
  {Arguello}, \citenamefont {Chockalingam}, \citenamefont {Rosenthal},
  \citenamefont {Zhao}, \citenamefont {Guti\'errez}, \citenamefont {Kang},
  \citenamefont {Chung}, \citenamefont {Fernandes}, \citenamefont {Jia},
  \citenamefont {Millis}, \citenamefont {Cava},\ and\ \citenamefont
  {Pasupathy}}]{Pasupathy_2014_PRB}%
  \BibitemOpen
  \bibfield  {author} {\bibinfo {author} {\bibfnamefont {C.~J.}\ \bibnamefont
  {Arguello}}, \bibinfo {author} {\bibfnamefont {S.~P.}\ \bibnamefont
  {Chockalingam}}, \bibinfo {author} {\bibfnamefont {E.~P.}\ \bibnamefont
  {Rosenthal}}, \bibinfo {author} {\bibfnamefont {L.}~\bibnamefont {Zhao}},
  \bibinfo {author} {\bibfnamefont {C.}~\bibnamefont {Guti\'errez}}, \bibinfo
  {author} {\bibfnamefont {J.~H.}\ \bibnamefont {Kang}}, \bibinfo {author}
  {\bibfnamefont {W.~C.}\ \bibnamefont {Chung}}, \bibinfo {author}
  {\bibfnamefont {R.~M.}\ \bibnamefont {Fernandes}}, \bibinfo {author}
  {\bibfnamefont {S.}~\bibnamefont {Jia}}, \bibinfo {author} {\bibfnamefont
  {A.~J.}\ \bibnamefont {Millis}}, \bibinfo {author} {\bibfnamefont {R.~J.}\
  \bibnamefont {Cava}}, \ and\ \bibinfo {author} {\bibfnamefont {A.~N.}\
  \bibnamefont {Pasupathy}},\ }\href {\doibase 10.1103/PhysRevB.89.235115}
  {\bibfield  {journal} {\bibinfo  {journal} {Physical Review B}\ }\textbf
  {\bibinfo {volume} {89}},\ \bibinfo {pages} {235115} (\bibinfo {year}
  {2014})}\BibitemShut {NoStop}%
\bibitem [{\citenamefont {Okamoto}\ and\ \citenamefont
  {Millis}(2015)}]{Millis_2015_PRB}%
  \BibitemOpen
  \bibfield  {author} {\bibinfo {author} {\bibfnamefont {J.-i.}\ \bibnamefont
  {Okamoto}}\ and\ \bibinfo {author} {\bibfnamefont {A.~J.}\ \bibnamefont
  {Millis}},\ }\href {\doibase 10.1103/PhysRevB.91.184204} {\bibfield
  {journal} {\bibinfo  {journal} {Physical Review B}\ }\textbf {\bibinfo
  {volume} {91}},\ \bibinfo {pages} {184204} (\bibinfo {year}
  {2015})}\BibitemShut {NoStop}%
\bibitem [{\citenamefont {Lian}\ \emph {et~al.}(2017)\citenamefont {Lian},
  \citenamefont {Wu}, \citenamefont {Xing}, \citenamefont {Wang},\ and\
  \citenamefont {Liu}}]{Liu_2017_PhysicaC}%
  \BibitemOpen
  \bibfield  {author} {\bibinfo {author} {\bibfnamefont {H.}~\bibnamefont
  {Lian}}, \bibinfo {author} {\bibfnamefont {Y.}~\bibnamefont {Wu}}, \bibinfo
  {author} {\bibfnamefont {H.}~\bibnamefont {Xing}}, \bibinfo {author}
  {\bibfnamefont {S.}~\bibnamefont {Wang}}, \ and\ \bibinfo {author}
  {\bibfnamefont {Y.}~\bibnamefont {Liu}},\ }\href {\doibase
  10.1016/j.physc.2017.05.003} {\bibfield  {journal} {\bibinfo  {journal}
  {Physica C: Superconductivity and its Applications}\ }\textbf {\bibinfo
  {volume} {538}},\ \bibinfo {pages} {27} (\bibinfo {year} {2017})}\BibitemShut
  {NoStop}%
\bibitem [{Sup()}]{SuppleRef}%
  \BibitemOpen
  \href@noop {} {}\bibinfo {note} {See Supplemental Material for more
  details.}\BibitemShut {Stop}%
\bibitem [{\citenamefont {Rice}\ \emph {et~al.}(1981)\citenamefont {Rice},
  \citenamefont {Whitehouse},\ and\ \citenamefont
  {Littlewood}}]{Littlewood_1981_PRB}%
  \BibitemOpen
  \bibfield  {author} {\bibinfo {author} {\bibfnamefont {T.~M.}\ \bibnamefont
  {Rice}}, \bibinfo {author} {\bibfnamefont {S.}~\bibnamefont {Whitehouse}}, \
  and\ \bibinfo {author} {\bibfnamefont {P.}~\bibnamefont {Littlewood}},\
  }\href {\doibase 10.1103/PhysRevB.24.2751} {\bibfield  {journal} {\bibinfo
  {journal} {Physical Review B}\ }\textbf {\bibinfo {volume} {24}},\ \bibinfo
  {pages} {2751} (\bibinfo {year} {1981})}\BibitemShut {NoStop}%
\bibitem [{\citenamefont {Crommie}\ \emph {et~al.}(1993)\citenamefont
  {Crommie}, \citenamefont {Lutz},\ and\ \citenamefont
  {Eigler}}]{Eigler_1993_Nature}%
  \BibitemOpen
  \bibfield  {author} {\bibinfo {author} {\bibfnamefont {M.~F.}\ \bibnamefont
  {Crommie}}, \bibinfo {author} {\bibfnamefont {C.~P.}\ \bibnamefont {Lutz}}, \
  and\ \bibinfo {author} {\bibfnamefont {D.~M.}\ \bibnamefont {Eigler}},\
  }\href {\doibase 10.1038/363524a0} {\bibfield  {journal} {\bibinfo  {journal}
  {Nature}\ }\textbf {\bibinfo {volume} {363}},\ \bibinfo {pages} {524}
  (\bibinfo {year} {1993})}\BibitemShut {NoStop}%
\bibitem [{\citenamefont {Brazovskii}\ \emph {et~al.}(2012)\citenamefont
  {Brazovskii}, \citenamefont {Brun}, \citenamefont {Wang},\ and\ \citenamefont
  {Monceau}}]{Monceau_2012_PRL}%
  \BibitemOpen
  \bibfield  {author} {\bibinfo {author} {\bibfnamefont {S.}~\bibnamefont
  {Brazovskii}}, \bibinfo {author} {\bibfnamefont {C.}~\bibnamefont {Brun}},
  \bibinfo {author} {\bibfnamefont {Z.-Z.}\ \bibnamefont {Wang}}, \ and\
  \bibinfo {author} {\bibfnamefont {P.}~\bibnamefont {Monceau}},\ }\href
  {\doibase 10.1103/PhysRevLett.108.096801} {\bibfield  {journal} {\bibinfo
  {journal} {Physical Review Letters}\ }\textbf {\bibinfo {volume} {108}},\
  \bibinfo {pages} {096801} (\bibinfo {year} {2012})}\BibitemShut {NoStop}%
\bibitem [{\citenamefont {Lee}\ and\ \citenamefont
  {Rice}(1979)}]{Rice_1979_PRB}%
  \BibitemOpen
  \bibfield  {author} {\bibinfo {author} {\bibfnamefont {P.~A.}\ \bibnamefont
  {Lee}}\ and\ \bibinfo {author} {\bibfnamefont {T.~M.}\ \bibnamefont {Rice}},\
  }\href {\doibase 10.1103/PhysRevB.19.3970} {\bibfield  {journal} {\bibinfo
  {journal} {Physical Review B}\ }\textbf {\bibinfo {volume} {19}},\ \bibinfo
  {pages} {3970} (\bibinfo {year} {1979})}\BibitemShut {NoStop}%
\bibitem [{\citenamefont {Johannes}\ and\ \citenamefont
  {Mazin}(2008)}]{Mazin_2008_PRB}%
  \BibitemOpen
  \bibfield  {author} {\bibinfo {author} {\bibfnamefont {M.~D.}\ \bibnamefont
  {Johannes}}\ and\ \bibinfo {author} {\bibfnamefont {I.~I.}\ \bibnamefont
  {Mazin}},\ }\href {\doibase 10.1103/PhysRevB.77.165135} {\bibfield  {journal}
  {\bibinfo  {journal} {Physical Review B}\ }\textbf {\bibinfo {volume} {77}},\
  \bibinfo {pages} {165135} (\bibinfo {year} {2008})}\BibitemShut {NoStop}%
\bibitem [{\citenamefont {Schackert}\ \emph {et~al.}(2015)\citenamefont
  {Schackert}, \citenamefont {M\"{a}rkl}, \citenamefont {Jandke}, \citenamefont
  {H\"{o}lzer}, \citenamefont {Ostanin}, \citenamefont {Gross}, \citenamefont
  {Ernst},\ and\ \citenamefont {Wulfhekel}}]{Wulfhekel_2015_PRL}%
  \BibitemOpen
  \bibfield  {author} {\bibinfo {author} {\bibfnamefont {M.}~\bibnamefont
  {Schackert}}, \bibinfo {author} {\bibfnamefont {T.}~\bibnamefont
  {M\"{a}rkl}}, \bibinfo {author} {\bibfnamefont {J.}~\bibnamefont {Jandke}},
  \bibinfo {author} {\bibfnamefont {M.}~\bibnamefont {H\"{o}lzer}}, \bibinfo
  {author} {\bibfnamefont {S.}~\bibnamefont {Ostanin}}, \bibinfo {author}
  {\bibfnamefont {E.~K.~U.}\ \bibnamefont {Gross}}, \bibinfo {author}
  {\bibfnamefont {A.}~\bibnamefont {Ernst}}, \ and\ \bibinfo {author}
  {\bibfnamefont {W.}~\bibnamefont {Wulfhekel}},\ }\href {\doibase
  10.1103/PhysRevLett.114.047002} {\bibfield  {journal} {\bibinfo  {journal}
  {Physical Review Letters}\ }\textbf {\bibinfo {volume} {114}},\ \bibinfo
  {pages} {047002} (\bibinfo {year} {2015})}\BibitemShut {NoStop}%
\bibitem [{\citenamefont {Minamitani}\ \emph {et~al.}(2017)\citenamefont
  {Minamitani}, \citenamefont {Arafune}, \citenamefont {Frederiksen},
  \citenamefont {Suzuki}, \citenamefont {Shahed}, \citenamefont {Kobayashi},
  \citenamefont {Endo}, \citenamefont {Fukidome}, \citenamefont {Watanabe},\
  and\ \citenamefont {Komeda}}]{Komeda_2017_PRB}%
  \BibitemOpen
  \bibfield  {author} {\bibinfo {author} {\bibfnamefont {E.}~\bibnamefont
  {Minamitani}}, \bibinfo {author} {\bibfnamefont {R.}~\bibnamefont {Arafune}},
  \bibinfo {author} {\bibfnamefont {T.}~\bibnamefont {Frederiksen}}, \bibinfo
  {author} {\bibfnamefont {T.}~\bibnamefont {Suzuki}}, \bibinfo {author}
  {\bibfnamefont {S.~M.~F.}\ \bibnamefont {Shahed}}, \bibinfo {author}
  {\bibfnamefont {T.}~\bibnamefont {Kobayashi}}, \bibinfo {author}
  {\bibfnamefont {N.}~\bibnamefont {Endo}}, \bibinfo {author} {\bibfnamefont
  {H.}~\bibnamefont {Fukidome}}, \bibinfo {author} {\bibfnamefont
  {S.}~\bibnamefont {Watanabe}}, \ and\ \bibinfo {author} {\bibfnamefont
  {T.}~\bibnamefont {Komeda}},\ }\href {\doibase 10.1103/PhysRevB.96.155431}
  {\bibfield  {journal} {\bibinfo  {journal} {Physical Review B}\ }\textbf
  {\bibinfo {volume} {96}},\ \bibinfo {pages} {155431} (\bibinfo {year}
  {2017})}\BibitemShut {NoStop}%
\bibitem [{\citenamefont {Zhang}\ \emph {et~al.}(2008)\citenamefont {Zhang},
  \citenamefont {Brar}, \citenamefont {Wang}, \citenamefont {Girit},
  \citenamefont {Yayon}, \citenamefont {Panlasigui}, \citenamefont {Zettl},\
  and\ \citenamefont {Crommie}}]{Crommie_2008_NP}%
  \BibitemOpen
  \bibfield  {author} {\bibinfo {author} {\bibfnamefont {Y.}~\bibnamefont
  {Zhang}}, \bibinfo {author} {\bibfnamefont {V.~W.}\ \bibnamefont {Brar}},
  \bibinfo {author} {\bibfnamefont {F.}~\bibnamefont {Wang}}, \bibinfo {author}
  {\bibfnamefont {C.}~\bibnamefont {Girit}}, \bibinfo {author} {\bibfnamefont
  {Y.}~\bibnamefont {Yayon}}, \bibinfo {author} {\bibfnamefont
  {M.}~\bibnamefont {Panlasigui}}, \bibinfo {author} {\bibfnamefont
  {A.}~\bibnamefont {Zettl}}, \ and\ \bibinfo {author} {\bibfnamefont {M.~F.}\
  \bibnamefont {Crommie}},\ }\href {\doibase 10.1038/nphys1022} {\bibfield
  {journal} {\bibinfo  {journal} {Nature Physics}\ }\textbf {\bibinfo {volume}
  {4}},\ \bibinfo {pages} {627} (\bibinfo {year} {2008})}\BibitemShut {NoStop}%
\bibitem [{\citenamefont {Stipe}\ \emph {et~al.}(1998)\citenamefont {Stipe},
  \citenamefont {Rezaei},\ and\ \citenamefont {Ho}}]{Ho_1998_Science}%
  \BibitemOpen
  \bibfield  {author} {\bibinfo {author} {\bibfnamefont {B.~C.}\ \bibnamefont
  {Stipe}}, \bibinfo {author} {\bibfnamefont {M.~A.}\ \bibnamefont {Rezaei}}, \
  and\ \bibinfo {author} {\bibfnamefont {W.}~\bibnamefont {Ho}},\ }\href
  {\doibase 10.1126/science.280.5370.1732} {\bibfield  {journal} {\bibinfo
  {journal} {Science}\ }\textbf {\bibinfo {volume} {280}},\ \bibinfo {pages}
  {1732} (\bibinfo {year} {1998})}\BibitemShut {NoStop}%
\bibitem [{\citenamefont {Heil}\ \emph {et~al.}(2018)\citenamefont {Heil},
  \citenamefont {Schlipf},\ and\ \citenamefont {Giustino}}]{Giustino_2018_PRB}%
  \BibitemOpen
  \bibfield  {author} {\bibinfo {author} {\bibfnamefont {C.}~\bibnamefont
  {Heil}}, \bibinfo {author} {\bibfnamefont {M.}~\bibnamefont {Schlipf}}, \
  and\ \bibinfo {author} {\bibfnamefont {F.}~\bibnamefont {Giustino}},\ }\href
  {\doibase 10.1103/PhysRevB.98.075120} {\bibfield  {journal} {\bibinfo
  {journal} {Physical Review B}\ }\textbf {\bibinfo {volume} {98}},\ \bibinfo
  {pages} {075120} (\bibinfo {year} {2018})}\BibitemShut {NoStop}%
\bibitem [{\citenamefont {Nishio}\ \emph {et~al.}(1994)\citenamefont {Nishio},
  \citenamefont {Shirai}, \citenamefont {Suzuki},\ and\ \citenamefont
  {Motizuki}}]{Motizuki_1994_JPSJ}%
  \BibitemOpen
  \bibfield  {author} {\bibinfo {author} {\bibfnamefont {Y.}~\bibnamefont
  {Nishio}}, \bibinfo {author} {\bibfnamefont {M.}~\bibnamefont {Shirai}},
  \bibinfo {author} {\bibfnamefont {N.}~\bibnamefont {Suzuki}}, \ and\ \bibinfo
  {author} {\bibfnamefont {K.}~\bibnamefont {Motizuki}},\ }\href {\doibase
  10.1143/JPSJ.63.156} {\bibfield  {journal} {\bibinfo  {journal} {Journal of
  the Physical Society of Japan}\ }\textbf {\bibinfo {volume} {63}},\ \bibinfo
  {pages} {156} (\bibinfo {year} {1994})}\BibitemShut {NoStop}%
\bibitem [{\citenamefont {Machida}\ \emph {et~al.}(2017)\citenamefont
  {Machida}, \citenamefont {Kohsaka}, \citenamefont {Iwaya}, \citenamefont
  {Arita}, \citenamefont {Hanaguri}, \citenamefont {Suzuki}, \citenamefont
  {Ochi},\ and\ \citenamefont {Iwasa}}]{Machida_2017_PRB}%
  \BibitemOpen
  \bibfield  {author} {\bibinfo {author} {\bibfnamefont {T.}~\bibnamefont
  {Machida}}, \bibinfo {author} {\bibfnamefont {Y.}~\bibnamefont {Kohsaka}},
  \bibinfo {author} {\bibfnamefont {K.}~\bibnamefont {Iwaya}}, \bibinfo
  {author} {\bibfnamefont {R.}~\bibnamefont {Arita}}, \bibinfo {author}
  {\bibfnamefont {T.}~\bibnamefont {Hanaguri}}, \bibinfo {author}
  {\bibfnamefont {R.}~\bibnamefont {Suzuki}}, \bibinfo {author} {\bibfnamefont
  {M.}~\bibnamefont {Ochi}}, \ and\ \bibinfo {author} {\bibfnamefont
  {Y.}~\bibnamefont {Iwasa}},\ }\href {\doibase 10.1103/PhysRevB.96.075206}
  {\bibfield  {journal} {\bibinfo  {journal} {Physical Review B}\ }\textbf
  {\bibinfo {volume} {96}},\ \bibinfo {pages} {075206} (\bibinfo {year}
  {2017})}\BibitemShut {NoStop}%
\bibitem [{\citenamefont {Sirica}\ \emph {et~al.}(2016)\citenamefont {Sirica},
  \citenamefont {Mo}, \citenamefont {Bondino}, \citenamefont {Pis},
  \citenamefont {Nappini}, \citenamefont {Vilmercati}, \citenamefont {Yi},
  \citenamefont {Gai}, \citenamefont {Snijders}, \citenamefont {Das},
  \citenamefont {Vobornik}, \citenamefont {Ghimire}, \citenamefont {Koehler},
  \citenamefont {Li}, \citenamefont {Sapkota}, \citenamefont {Parker},
  \citenamefont {Mandrus},\ and\ \citenamefont {Mannella}}]{Sirica_2016_PRB}%
  \BibitemOpen
  \bibfield  {author} {\bibinfo {author} {\bibfnamefont {N.}~\bibnamefont
  {Sirica}}, \bibinfo {author} {\bibfnamefont {S.-K.}\ \bibnamefont {Mo}},
  \bibinfo {author} {\bibfnamefont {F.}~\bibnamefont {Bondino}}, \bibinfo
  {author} {\bibfnamefont {I.}~\bibnamefont {Pis}}, \bibinfo {author}
  {\bibfnamefont {S.}~\bibnamefont {Nappini}}, \bibinfo {author} {\bibfnamefont
  {P.}~\bibnamefont {Vilmercati}}, \bibinfo {author} {\bibfnamefont
  {J.}~\bibnamefont {Yi}}, \bibinfo {author} {\bibfnamefont {Z.}~\bibnamefont
  {Gai}}, \bibinfo {author} {\bibfnamefont {P.~C.}\ \bibnamefont {Snijders}},
  \bibinfo {author} {\bibfnamefont {P.~K.}\ \bibnamefont {Das}}, \bibinfo
  {author} {\bibfnamefont {I.}~\bibnamefont {Vobornik}}, \bibinfo {author}
  {\bibfnamefont {N.}~\bibnamefont {Ghimire}}, \bibinfo {author} {\bibfnamefont
  {M.~R.}\ \bibnamefont {Koehler}}, \bibinfo {author} {\bibfnamefont
  {L.}~\bibnamefont {Li}}, \bibinfo {author} {\bibfnamefont {D.}~\bibnamefont
  {Sapkota}}, \bibinfo {author} {\bibfnamefont {D.~S.}\ \bibnamefont {Parker}},
  \bibinfo {author} {\bibfnamefont {D.~G.}\ \bibnamefont {Mandrus}}, \ and\
  \bibinfo {author} {\bibfnamefont {N.}~\bibnamefont {Mannella}},\ }\href
  {\doibase 10.1103/PhysRevB.94.075141} {\bibfield  {journal} {\bibinfo
  {journal} {Physical Review B}\ }\textbf {\bibinfo {volume} {94}},\ \bibinfo
  {pages} {075141} (\bibinfo {year} {2016})}\BibitemShut {NoStop}%
\bibitem [{\citenamefont {Hou}\ \emph {et~al.}(2020)\citenamefont {Hou},
  \citenamefont {Zhang}, \citenamefont {Tu}, \citenamefont {Gu}, \citenamefont
  {Zhang}, \citenamefont {Gong}, \citenamefont {Tu}, \citenamefont {Wang},
  \citenamefont {Lv}, \citenamefont {Weng}, \citenamefont {Ren}, \citenamefont
  {Chen}, \citenamefont {Zhu}, \citenamefont {Hao},\ and\ \citenamefont
  {Shan}}]{IETS_PRL}%
  \BibitemOpen
  \bibfield  {author} {\bibinfo {author} {\bibfnamefont {X.-Y.}\ \bibnamefont
  {Hou}}, \bibinfo {author} {\bibfnamefont {F.}~\bibnamefont {Zhang}}, \bibinfo
  {author} {\bibfnamefont {X.-H.}\ \bibnamefont {Tu}}, \bibinfo {author}
  {\bibfnamefont {Y.-D.}\ \bibnamefont {Gu}}, \bibinfo {author} {\bibfnamefont
  {M.-D.}\ \bibnamefont {Zhang}}, \bibinfo {author} {\bibfnamefont
  {J.}~\bibnamefont {Gong}}, \bibinfo {author} {\bibfnamefont {Y.-B.}\
  \bibnamefont {Tu}}, \bibinfo {author} {\bibfnamefont {B.-T.}\ \bibnamefont
  {Wang}}, \bibinfo {author} {\bibfnamefont {W.-G.}\ \bibnamefont {Lv}},
  \bibinfo {author} {\bibfnamefont {H.-M.}\ \bibnamefont {Weng}}, \bibinfo
  {author} {\bibfnamefont {Z.-A.}\ \bibnamefont {Ren}}, \bibinfo {author}
  {\bibfnamefont {G.-F.}\ \bibnamefont {Chen}}, \bibinfo {author}
  {\bibfnamefont {X.-D.}\ \bibnamefont {Zhu}}, \bibinfo {author} {\bibfnamefont
  {N.}~\bibnamefont {Hao}}, \ and\ \bibinfo {author} {\bibfnamefont
  {L.}~\bibnamefont {Shan}},\ }\href {\doibase 10.1103/PhysRevLett.124.106403}
  {\bibfield  {journal} {\bibinfo  {journal} {Phys. Rev. Lett.}\ }\textbf
  {\bibinfo {volume} {124}},\ \bibinfo {pages} {106403} (\bibinfo {year}
  {2020})}\BibitemShut {NoStop}%
\end{thebibliography}

\end{document}